\documentclass{article}
\usepackage{spconf,amsmath,graphicx,hyperref}

\usepackage{spconf,amsmath,graphicx}
\usepackage{multirow}
\usepackage{booktabs}
\usepackage{url}
\usepackage{subcaption}
\usepackage{siunitx}
\usepackage{pgfplots}
\usepackage{algorithm}
\usepackage{algorithmic} 
\usepackage{tikz}
\usepackage{amsmath}  
\usetikzlibrary{shapes, arrows, positioning}
\usepackage{fontawesome5} 
\usepackage{bbm} 
\usepackage{fontawesome5} 
\pgfplotsset{compat=1.18}
\title{CLEAR: Online Speech Content Leakage Estimation through Cross-ASR Disagreement}

\name{Bhawana Chhaglani$^*$, Tanvi Kandepuneni$^*$, Jeremy Gummeson, Prashant Shenoy\thanks{$^{*}$These authors contributed equally to this work.}}
\address{University of Massachusetts Amherst}

\begin{document}
\ninept
\maketitle
\begin{abstract}
Signal-level speech privacy mechanisms suppress linguistic content while preserving acoustic information needed by downstream sensing applications. However, their privacy settings are typically evaluated/selected offline and remain fixed during deployment, even though speech-content leakage can vary substantially across utterances and speakers. Adapting privacy protection at runtime requires estimating how much speech remains recoverable, but conventional measures such as WER or PER require ground-truth transcripts and therefore cannot be computed online.
We present CLEAR, a reference-free approach for estimating speech-content leakage at runtime using disagreement among heterogeneous ASR systems. Our key insight is that independently trained ASRs exhibit consistent behavior when linguistic content remains recoverable and increasingly disagree as privacy transformations obscure speech. Using configurable speech-suppression mechanism, we show that cross-ASR disagreement closely tracks transcript-grounded leakage across privacy operating points, achieving a correlation of 0.8. 
We further characterize the latency-accuracy trade-off of heterogeneous ASR subsets and use their hypotheses to identify potentially exposed words.
These capabilities enable privacy to be treated as a runtime property rather than a fixed configuration: CLEAR can communicate residual speech exposure to users and provide feedback for dynamically adjusting privacy aggressiveness while retaining acoustic utility for downstream sensing tasks.
\end{abstract}

\vspace{-0.3cm}

\section{Introduction}
\vspace{-0.2cm}
Microphones embedded in smartphones, wearables, and ambient devices enable a growing range of applications, including activity recognition \cite{mollyn2022samosa,chhaglani2024aerosense}, health monitoring \cite{sun2015symdetector,chhaglani2026poster}, and environmental sensing \cite{chhaglani2022flowsense}. However, these same audio streams also capture sensitive linguistic content, inadvertently exposing personal health conditions, financial details, and other private conversations \cite{chhaglani2024towards,chhaglani2025featuresense}. Prior work has therefore proposed signal-level mechanisms that suppress speech while retaining acoustic information useful for downstream sensing. For example, 
Kirigami \cite{boovaraghavan2024kirigami} selectively detects and removes speech regions while preserving non-speech audio for activity recognition. Other techniques include low pass filtering \cite{iravantchi2021privacymic,liu2024private},subsampling \cite{mollyn2022samosa}, selectively sampling \cite{laput2017synthetic}, etc. Importantly, each mechanism exposes a parameter that control the aggressiveness of the filtering and, consequently, its privacy–utility trade-off. 

Selecting these parameters and evaluating these techniques is challenging because speech-content privacy is not fixed for a given configuration. Privacy settings are typically selected offline based on aggregate evaluation over a dataset and remain fixed during deployment. However, privacy is input-dependent and can vary substantially across utterances and speakers even under the same transformation and parameter setting. We illustrate this using our Common Voice evaluation subset \cite{ardila2020common}, which contains short speech utterances from speakers with diverse genders, ages, and accents. With Kirigami at a fixed threshold of $\tau=0.5$, adversarial PER varies from 0.45 to 1.0 across utterances. We also observe substantial speaker-level differences: S1, a male speaker with Canadian English, has a mean PER of $0.61\pm0.30$, whereas S2, a male non-native speaker with German English, has a mean PER of $0.94\pm0.17$ at the same operating point. These results demonstrate both within- and across-speaker variability in the privacy provided by a fixed configuration.
A fixed configuration may therefore leave some speech substantially more recoverable while unnecessarily suppressing useful acoustic information for others. Rather than assigning a fixed privacy level to an operating point, a system should ideally assess the leakage of the audio currently being processed and adapt its privacy aggressiveness accordingly.

Such adaptation requires measuring speech-content leakage at runtime, which existing evaluation methods cannot readily provide. Prior work has highlighted the importance and difficulty of quantifying residual linguistic information in privacy-preserving audio~\cite{williams2023new,williams2021revisiting}. Existing evaluations typically apply an ASR attack and compare its output against the ground-truth transcript using word, character, or phoneme error rate (WER/CER/PER)~\cite{liu2024private,boovaraghavan2024kirigami}. While effective offline, these metrics require knowledge of what was originally spoken - a reference that is unavailable, and undesirable to retain, during privacy-preserving deployment. This prevents a system from determining its current privacy exposure, communicating that exposure to users, or deciding when stronger protection is needed. What is therefore missing is a \emph{reference-free runtime signal} of residual speech-content leakage.

In this work, we investigate whether agreement among heterogeneous ASR systems can provide this missing signal. Our key intuition is that independently trained ASRs, which differ in architecture, training data, and decoding strategies, are likely to make different errors when speech content is poorly recoverable. If they nevertheless recover similar linguistic content from privacy-transformed audio, that agreement provides evidence that the content remains exposed; as recoverability decreases, their hypotheses should increasingly diverge. 
Based on this insight, we introduce \textit{Content Leakage Estimation through ASR disagreement (CLEAR)}, a reference-free measure of runtime speech-content leakage. 
CLEAR processes only the privacy-transformed audio and requires neither the original audio nor its ground-truth transcript. Beyond producing a leakage estimate, the ASR hypotheses expose words that are consistently recovered across recognizers, allowing a system to communicate \emph{what content may be leaking} rather than only reporting an abstract privacy score. Most importantly, CLEAR provides a feedback signal for adaptive privacy: when estimated leakage increases, a system can increase the aggressiveness of the privacy transformation; when leakage is sufficiently low, it can use a less aggressive configuration to retain additional acoustic information for downstream sensing tasks.

We evaluate CLEAR using Kirigami~\cite{boovaraghavan2024kirigami} on Common Voice dataset
across 13 privacy operating points. CLEAR strongly
tracks transcript-grounded speech leakage from a held-out strong ASR adversary,
achieving a Spearman correlation of $\rho=0.80$ across
utterance-configuration pairs 
and $\rho=0.98$ across operating-point
averages. 
CLEAR can additionally identify
potentially exposed words with 61.3\% correctness, enabling interpretable runtime privacy feedback. We also show that CLEAR is better than using a single ASR confidence score and works across different privacy transformations. Lastly, we find an optimal ASR ensemble that estimates speech leakage with low latency (RTF = 0.15).
\vspace{-0.2cm}
\section{Background and Motivation}
\label{sec:background}
\vspace{-0.1cm}
\subsection{Speech Content Privacy}
\vspace{-0.1cm}
Speech privacy encompasses speaker identity~\cite{panariello2024voiceprivacy,tran2023speech}, paralinguistic attributes~\cite{chhaglani2025featuresense}, and linguistic content~\cite{boovaraghavan2024kirigami,williams2023new,williams2023privacy}. We focus on speech-content privacy, i.e., limiting recovery of what was spoken. Prior work highlights that sensitive linguistic content can remain exposed even when speaker identity is concealed~\cite{williams2023new,williams2021revisiting}. Metrics such as WER, PER, and Masked Error Rate (MER)~\cite{williams2021revisiting} quantify such leakage by comparing recovered speech against a ground-truth transcript, making them suitable for offline evaluation but unavailable during deployment.
Signal-level privacy mechanisms suppress speech while retaining cues useful for sensing, using techniques such as low-pass filtering~\cite{iravantchi2021privacymic,liu2024private}, subsampling~\cite{mollyn2022samosa}, selective sampling~\cite{laput2017synthetic}, and speech suppression or obfuscation~\cite{xia2020pams,boovaraghavan2024kirigami,nicolaou2026removing}. CLEAR addresses the runtime evaluation gap of these techniques by estimating residual speech content  without requiring the ground-truth transcript.
\vspace{-0.6cm}
\subsection{Reference-Free Speech Intelligibility Prediction}
\vspace{-0.2cm}
Speech intelligibility is traditionally evaluated through human studies or reference-based metrics such as STOI and PESQ~\cite{kondo2012subjective,weismer2008speech}. More recent non-intrusive approaches estimate human intelligibility directly from degraded speech using ASR uncertainty or learned representations~\cite{tu2022unsupervised,karbasi2020non,zezario2020stoi, mogridge2024non}. These methods primarily target \emph{human perceptual intelligibility}; in contrast, we measure whether linguistic content remains recoverable by modern ASR systems, representing a practical machine adversary.
CLEAR leverages ASR behavior but, rather than relying on a single recognizer, measures disagreement in the content recovered by heterogeneous ASRs.
\vspace{-0.3cm}
\subsection{Runtime Privacy Awareness: Opportunities}
\vspace{-0.2cm}
A reference-free estimate of speech-content leakage enables capabilities beyond offline evaluation. First, it can provide \emph{user-facing privacy awareness} by reporting residual leakage and identifying words consistently recovered across ASRs. Second, it can provide \emph{privacy-control feedback}: when leakage is high, the system can increase privacy aggressiveness, and when leakage is low, relax it to preserve useful acoustic information. This enables an \emph{adaptive privacy mode} that continuously limits speech-content exposure while retaining non-speech audio for applications such as emergency-sound detection on Alexa \cite{amazon_alexa_aed} and health sensing. Thus, privacy mechanisms can move from static offline configurations toward monitoring, communicating, and adapting privacy during deployment.
\vspace{-0.2cm}

\section{Method}
\label{sec:method}
    \vspace{-0.4cm}
\begin{figure}[t]
    \centering
    \includegraphics[width=0.99\columnwidth]{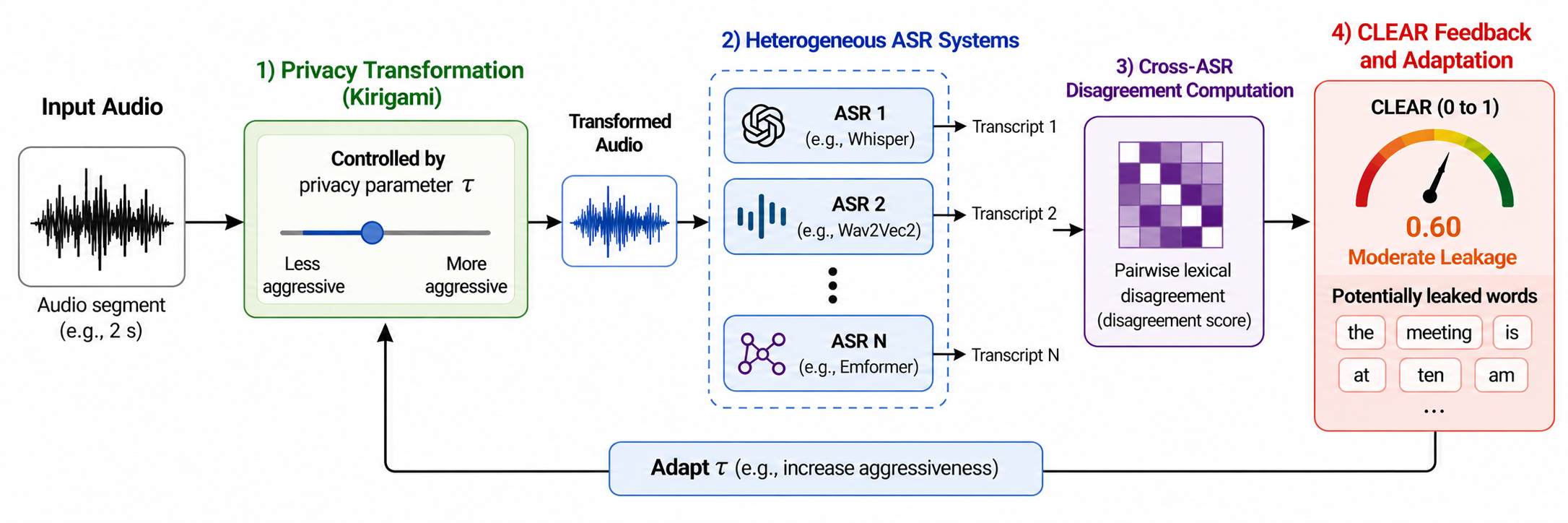}
    \caption{\textsc{CLEAR}: Privacy-transformed audio is processed by heterogeneous ASR systems, and disagreement among their hypotheses is used to estimate speech-content leakage and leaked words.} 
    \label{fig:clear_pipeline}
    \vspace{-0.5cm}
\end{figure}
\noindent\textbf{Threat Model}: We consider an audio sensing system that applies a signal-level privacy mechanism \(T\) before storage, transmission, or downstream processing. An adversary obtains the transformed signal \(x'=T(x)\) and attempts to recover linguistic content using a modern ASR attackers. Our objective is not to conceal speaker identity, emotion, or other paralinguistic attributes; CLEAR specifically measures residual linguistic content recoverability.
During offline evaluation, ground-truth transcripts are available to quantify adversarial recovery. During deployment, CLEAR receives only \(x'\).

\noindent\textbf{Cross-ASR Disagreement}
\label{sec:cross_asr}
CLEAR is based on the intuition that independently trained ASR systems should recover similar content when speech remains intelligible after a privacy transformation as shown in Figure \ref{fig:clear_pipeline}. Conversely, as the transformation obscures linguistic information, the ASRs are increasingly likely to produce different hypotheses. We therefore use \emph{cross-ASR disagreement} as a reference-free signal of speech-content privacy.
Given privacy-transformed audio \(x'\), we process it using a set of \(N\) heterogeneous ASR systems,
    $\mathcal{A} = \{A_1,A_2,\ldots,A_N\},$
where each recognizer produces a hypothesis
 $ h_i = A_i(x')$, 
 $\qquad i \in \{1,\ldots,N\}.$
We normalize each hypothesis by converting it to lowercase, removing punctuation, and normalizing whitespace. Let \(\tilde{h}_i\) denote the resulting normalized transcript.
For each pair of recognizers \(A_i\) and \(A_j\), we compute transcript similarity using sequence matching,
    $s_{ij} =
    \mathrm{SeqSim}(\tilde{h}_i,\tilde{h}_j),$
where \(s_{ij}\in[0,1]\), with larger values indicating more similar ASR hypotheses. Pairwise disagreement is then defined as $d_{ij}=1-s_{ij}.$
Empty ASR hypotheses require special treatment. If exactly one recognizer produces an empty hypothesis, we assign \(d_{ij}=1\), since one recognizer recovered content while the other did not. If both hypotheses are empty, we still assign \(d_{ij}=1\) even though there is agreement: but it indicates zero privacy leakage. 
We also separately track the fraction of ASRs producing empty hypotheses,
which captures ASR recovery failure independently of cross-ASR disagreement.
We define \textsc{CLEAR} as the mean disagreement across all ASR pairs (as shown in Algorithm \ref{alg:clear}).
A higher \textsc{CLEAR} score indicates greater disagreement among heterogeneous ASRs and therefore lower estimated speech-content recoverability, whereas a lower score indicates that multiple ASRs recover similar content and hence suggests greater potential leakage. 

\begin{algorithm}[t]
\footnotesize
\caption{\textsc{CLEAR}: Reference-Free Cross-ASR Disagreement}
\label{alg:clear}
\begin{algorithmic}[1]
\STATE \textbf{Input:} transformed audio $x'$, ASRs $\mathcal{A}=\{A_1,\ldots,A_N\}$
\STATE \textbf{Output:} CLEAR score $c\in[0,1]$, empty rate $e\in[0,1]$, leaked words $L$

\FOR{$i=1,\ldots,N$}
    \STATE $h_i \leftarrow \mathrm{Normalize}(A_i(x'))$
\ENDFOR

\STATE $D \leftarrow 0$, $L \leftarrow \emptyset$

\FOR{each pair $(i,j),\,i<j$}
    \IF{$h_i=\emptyset$ \AND $h_j=\emptyset$}
        \STATE $d_{ij} \leftarrow 1$
    \ELSIF{$h_i=\emptyset$ \OR $h_j=\emptyset$}
        \STATE $d_{ij} \leftarrow 1$
    \ELSE
        \STATE $d_{ij} \leftarrow 1-\mathrm{SeqSim}(h_i,h_j)$
        \STATE $L \leftarrow L \cup \mathrm{MatchingWords}(h_i,h_j)$
    \ENDIF
    \STATE $D \leftarrow D+d_{ij}$
\ENDFOR

\STATE $c \leftarrow \frac{2D}{N(N-1)}$
\STATE $e \leftarrow \frac{1}{N}\sum_{i=1}^{N}\mathbbm{1}[h_i=\emptyset]$

\RETURN $c,e,L$
\end{algorithmic}
\vspace{-0.1cm}
\end{algorithm}
\normalsize

\noindent\textbf{Potentially Leaked Words.}
In addition to estimating overall speech-content recoverability, CLEAR identifies words that may remain exposed. We consider a word potentially leaked if it is recovered by at least two heterogeneous ASRs. Specifically, $\mathrm{MatchingWords}(h_i,h_j)$ returns the words shared between a pair of normalized ASR hypotheses, and we take their union across all ASR pairs to obtain the leaked-word set $L$. These words can be communicated to the user as a privacy alert, similar to security alerts on personal devices, informing users not only that potential leakage was detected but also which words may have been exposed.

\noindent\textbf{Privacy-control feedback.}
CLEAR can serve as a runtime feedback signal for adaptive privacy control. When CLEAR falls below a desired privacy target, indicating greater cross-ASR agreement and potential leakage, the system can increase the aggressiveness of the privacy transformation and recompute CLEAR. This feedback loop can continue until the desired privacy level is reached, enabling privacy adaptation rather than relying on a fixed privacy configuration.

\vspace{-0.3cm}
\section{Evaluation}
\vspace{-0.3cm}
\label{sec:evaluation}
\subsection{Experimental Setup}
\label{sec:setup}

\noindent\textbf{Dataset.}
We curate a subset of 100 speech utterances (\~2 seconds each) from Mozilla Common Voice~\cite{ardila2020common}, a large-scale speech corpus containing recordings from diverse speakers with corresponding reference transcripts. The transcripts are used \emph{only} to establish ground-truth leakage during offline evaluation and are never provided to CLEAR at runtime.

\noindent\textbf{Privacy Transformations.}
We primarily evaluate CLEAR using Kirigami~\cite{boovaraghavan2024kirigami}, which detects and selectively suppresses speech regions while retaining non-speech acoustic information. Kirigami exposes a threshold ($\tau$) that controls the privacy--utility trade-off. We sweep 13 thresholds from 0.3 to 0.9, producing operating points with varying levels of residual speech recoverability. This allows us to evaluate whether CLEAR tracks changes in leakage as transformation aggressiveness varies. To evaluate generalizability beyond Kirigami, we additionally evaluate CLEAR with low-pass filtering (LPF)~\cite{iravantchi2021privacymic,liu2024private}, sweeping cutoff frequencies of 300, 500, and 700,Hz.

\noindent\textbf{ASR Models.}
We use six SOTA heterogeneous ASR systems to compute CLEAR: Wav2Vec2-base-960h, whisper-small-en, Citrinet-1024-$\gamma$-0.25, speechbrain-crdnn-rnnlm, emformer-rnnt-librispeech, and Parakeet-TDT-0.6B-v3. These models span substantially different recognition architectures and training paradigms: for exmaple, wav2Vec2 uses self-supervised speech representations followed by CTC-based recognition; Citrinet is a convolutional CTC acoustic model; and Parakeet uses a FastConformer encoder with a token-and-duration transducer (TDT) decoder. This heterogeneity is intentional: CLEAR relies on agreement or disagreement among recognizers with different inductive biases rather than repeated predictions from closely related ASR models.

\noindent\textbf{Metrics.}
Although CLEAR itself is reference-free, we use reference transcripts to quantify actual content recoverability for evaluation purpose. We primarily use PER, since privacy transformations can produce lexical errors while still preserving substantial phonetic information about what was spoken. This PER is computed using hubert-large-ls960-ft and the groundtruth transcript. We convert both the reference transcript and adversarial ASR hypothesis to a common phoneme representation and compute their normalized edit distance. Higher PER therefore indicates lower speech-content recoverability and stronger privacy.
We evaluate the relationship between CLEAR and transcript-grounded PER using both Spearman rank correlation ($\rho$) and Pearson correlation ($r$). Spearman correlation measures whether CLEAR correctly tracks the ordering of privacy levels without assuming a linear relationship, while Pearson correlation captures linear association. We report correlations across individual utterance-configuration pairs, across privacy operating-point averages, and within individual utterances as the privacy parameter is varied. 
We measure the inference latency of each ASR and the resulting CLEAR computation in milliseconds on a Google Colab NVIDIA T4 GPU. 

\vspace{-0.3cm}

\subsection{Results}
\vspace{-0.1cm}
\noindent\textbf{Selecting a held-out ASR adversary.}
To obtain a good proxy of speech content leakage (transcript-grounded measure), we first identify the strongest ASR adversary among the models considered. 
For this, we apply various privacy transformations (low-pass filtering at 300 Hz \cite{iravantchi2021privacymic}, subsampling at 1 kHz \cite{mollyn2022samosa}, selective sampling 10 samples/s \cite{laput2017synthetic}, speech suppression 0.3 and 0.5 \cite{boovaraghavan2024kirigami}) on 35 audio files to understand which ASR is strong enough to recover speech even under aggressive privacy filtering. 
Table~\ref{tab:asr_adversary} reports the mean PER of each ASR across the evaluated privacy transformations. HuBERT achieves the lowest mean PER (0.936), indicating the strongest overall recovery of linguistic content from privacy-transformed audio. We therefore hold out HuBERT from the ASRs used to compute CLEAR and use its PER against the ground-truth transcript as our primary measure of actual speech-content recoverability. This separation ensures that CLEAR is evaluated against an independent ASR adversary rather than one of the recognizers used to construct the disagreement score.
\begin{table}[t]
\centering
\begin{minipage}[t]{0.48\columnwidth}
\centering
\caption{\small Mean PER of candidate ASR adversaries across the aggressive privacy transformations.}
\label{tab:asr_adversary}
\footnotesize
\setlength{\tabcolsep}{5pt}
\begin{tabular}{lc}
\toprule
\textbf{ASR} & \textbf{Mean PER} \\
\midrule
HuBERT      & \textbf{0.936} \\
Emformer    & 0.955 \\
Citrinet    & 0.969 \\
Wav2Vec2    & 0.976 \\
Whisper     & 1.005 \\
Parakeet    & 1.043 \\
SpeechBrain & 1.082 \\
\bottomrule
\end{tabular}
\end{minipage}
\hfill
\begin{minipage}[t]{0.48\columnwidth}
\centering
\caption{\small Performance of CLEAR calibration methods for predicting adversarial PER.}
\label{tab:calibration}
\footnotesize
\setlength{\tabcolsep}{6pt}
\begin{tabular}{lc}
\toprule
\textbf{Method} & \textbf{MAE $\downarrow$} \\
\midrule
Isotonic calibration & \textbf{0.162} \\
Linear calibration   & 0.172 \\
Mean-PER baseline    & 0.316 \\
\bottomrule
\end{tabular}
\end{minipage}

\vspace{-0.2cm}
\end{table}
\normalsize

\begin{figure}
    \centering
    \includegraphics[width=0.9\linewidth]{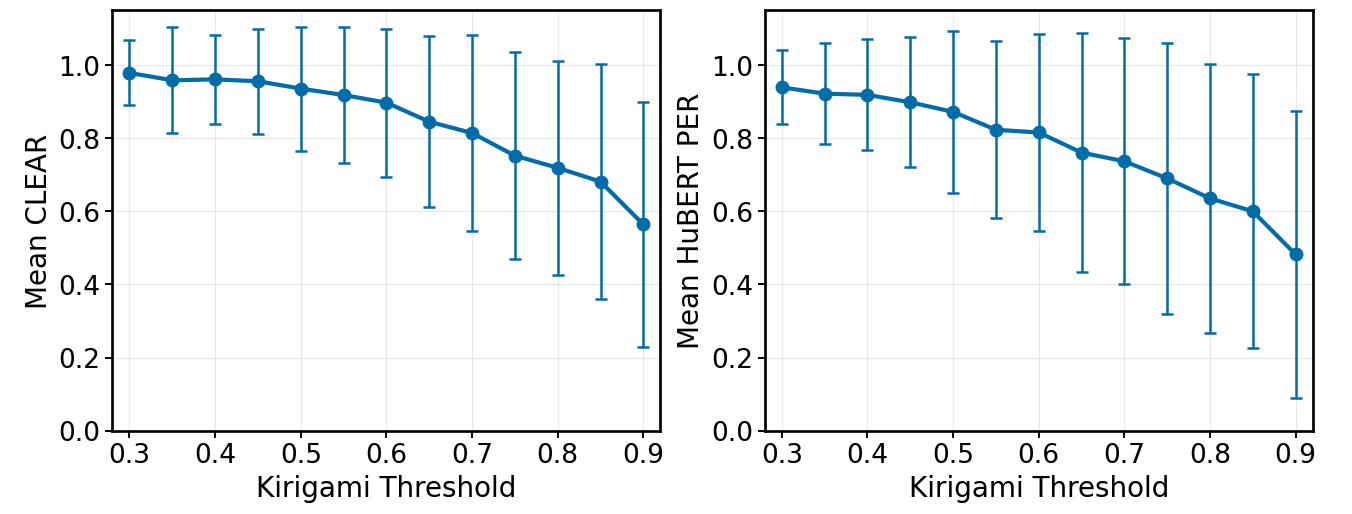}
    \caption{
Mean CLEAR disagreement (left) and held-out HuBERT PER (right) across 13 Kirigami operating points and 100 utterances. Both follow similar trends as privacy aggressiveness varies.}
    \label{fig:clear_per}
    \vspace{-0.5cm}
\end{figure}



\noindent\textbf{CLEAR tracks speech-content recoverability.}
We evaluate whether CLEAR provides a reliable reference-free indicator of speech-content privacy by comparing it with transcript-grounded PER from the held-out HuBERT adversary. Across all 100 utterances and 13 Kirigami operating points ($n=1300$), CLEAR exhibits a strong positive association with held-out PER (Spearman $\rho=0.8$). Spearman correlation is computed over the 1,300 individual utterance--configuration pairs and measures whether higher CLEAR disagreement consistently corresponds to higher PER, independent of the exact scale of the two metrics. The strong correlations indicate that when the CLEAR ensemble increasingly disagrees about the recovered content, speech recoverability reduces.
Figure~\ref{fig:clear_per} further examines whether CLEAR tracks changes in privacy as Kirigami's operating parameter is varied. We average CLEAR and held-out PER across the 100 utterances separately at each of the 13 operating points and correlate these 13 operating-point means. CLEAR closely tracks the resulting privacy continuum (Spearman $\rho=0.98$): operating points with greater cross-ASR disagreement also produce higher PER for the held-out adversary. Together, the utterance-level and operating-point analyses show that CLEAR captures both instance-level variation in speech recoverability and systematic changes in privacy induced by adjusting Kirigami's aggressiveness, without requiring access to the reference transcript. At the individual-utterance level, CLEAR also tracks changes in recoverability as the privacy parameter varies, achieving a median within-utterance Spearman correlation of $\rho=0.82$ across the 13 Kirigami operating points.
We additionally examine ASR failure through the fraction of empty transcripts. Empty-output rate increases with speech-content privacy and is itself correlated with held-out PER ($\rho=0.534$). Similarly, CLEAR shows similar correlation ($\rho=0.602$). 

\noindent\textbf{Calibrating CLEAR to adversarial recoverability.}
While CLEAR provides a reference-free measure of relative speech-content leakage, its raw score does not directly correspond to adversarial recoverability. We therefore learn a calibration function that maps CLEAR scores to estimated PER. We evaluate both linear and isotonic regression, the latter capturing the expected monotonic relationship between CLEAR and PER without assuming linearity. To avoid leakage across operating points of the same utterance, we evaluate calibration using 5-fold cross-validation grouped by utterance, such that all 13 configurations of a held-out utterance remain exclusively in the test fold. As shown in Table~\ref{tab:calibration}, isotonic calibration achieves the lowest prediction error (MAE $=0.162$), slightly outperforming linear calibration (MAE $=0.172$) and substantially improving over a mean-PER baseline (MAE $=0.316$). This calibration enables a deployment to translate runtime CLEAR scores into estimated PER and compare them against an application-specific privacy target. 
As an illustrative deployment target, classifying whether adversarial PER is at least 0.7 achieves 82\% accuracy. Thus, rather than assigning universal notions of safe or unsafe to raw CLEAR values, a deployment can calibrate CLEAR offline against an application-selected recoverability target and use the resulting mapping for runtime privacy-control feedback.

\noindent\textbf{Comparison with ASR confidence.}
Alternatively, the confidence of a single ASR as a runtime indicator of speech recoverability can be used. However, ASR confidence is model-specific and does not directly measure residual speech content. On the same evaluation subset, Wav2Vec2 confidence is negatively correlated with held-out HuBERT PER, as expected, but achieves only moderate association (Spearman $\rho=-0.626$). So, we cannot use single ASR confidence for speech privacy leakage evaluation.

\begin{table}[t]
\centering
\caption{\textbf{Identification of potentially leaked words.}
Word recovery is evaluated over 30 clips containing 74 words at three Kirigami
operating points.}
\label{tab:word_recovery}
\footnotesize
\setlength{\tabcolsep}{5pt}
\begin{tabular}{c c c c}
\toprule
\textbf{Kirigami} &
\textbf{Common} &
\textbf{Recovered} &
\textbf{Recovery} \\
\textbf{Threshold} &
\textbf{Words} &
\textbf{Words} &
\textbf{Rate} \\
\midrule
0.3 & 17 &  8 & 10.8\% \\
0.5 & 28 & 17 & 23.0\% \\
0.7 & 42 & 32 & 43.2\% \\
\bottomrule
\end{tabular}
\vspace{-0.15cm}
\end{table}
\normalsize

\noindent\textbf{CLEAR identifies potentially leaked words.}
Beyond estimating overall speech recoverability, CLEAR can identify
specific words that may remain exposed. We flag a word as
potentially leaked when it is recovered by at least two
heterogeneous ASRs. We evaluate these alerts over 30 utterances
containing 74 reference words at three Kirigami operating points.
As speech becomes increasingly recoverable, the fraction of
reference words correctly identified by CLEAR increases from
10.8\% at threshold 0.3 to 23.0\% at 0.5 and 43.2\% at 0.7
(Table~\ref{tab:word_recovery}). At threshold 0.7, CLEAR identifies
32 of the 74 spoken words without accessing the reference
transcript. This demonstrates that cross-ASR agreement can provide
users with interpretable feedback about \emph{what} linguistic
content may remain exposed, rather than only reporting a scalar
privacy score.
One interesting observation is that in most cases (61.3\% on average), the common words are infact the leaked words, which indicates that ASRs agreement is a useful signal for leakage.

\noindent\textbf{CLEAR generalizes beyond Kirigami.}
To examine whether CLEAR captures speech recoverability rather
than properties specific to Kirigami, we additionally evaluate
low-pass filtering at cutoff frequencies of 300, 500, and
700\,Hz over 10 utterances. 
We observe that CLEAR follows the same trend as held-out
HuBERT PER: mean CLEAR decreases from 0.953 at 300\,Hz to
0.594 at 700\,Hz, while mean PER decreases from 0.827 to
0.576. Across the three operating-point means, CLEAR and PER
have correlation of 0.832.
This shows that CLEAR's disagreement signal is not specific to Kirigami
and can track changes in speech recoverability under a different
signal-level privacy transformation. Future work will explore evaluating CLEAR on other privacy techniques.

\noindent\textbf{Accuracy--latency trade-off.}
The primary computational cost of CLEAR arises from ASR inference:
the disagreement metric itself requires less than 0.15\,ms even
for six ASRs. 
We evaluate all possible ASR subsets and report the highest-correlation ensemble
for each size in Table~\ref{tab:ensemble_tradeoff}. A three-ASR ensemble
achieves $\rho=0.807$ with only 0.30\,s sequential inference latency for
a ${\sim}$2-s clip (RTF $=0.15$). Adding a fourth ASR provides negligible
improvement ($\rho=0.81$) while increasing latency to 1.34\,s.
Thus, three heterogeneous ASRs provide a favorable accuracy--latency
trade-off for runtime CLEAR estimation.

\begin{table}[t]
\centering
\caption{\textbf{CLEAR accuracy--latency trade-off.}
Latency is measured on an NVIDIA Tesla T4 GPU for ${\sim}$2-s clips.}
\label{tab:ensemble_tradeoff}
\footnotesize
\setlength{\tabcolsep}{3.5pt}
\begin{tabular}{c l c c}
\toprule
\textbf{\#} & \textbf{Best ASR subset} &
$\boldsymbol{\rho}$ & \textbf{Latency (s)} \\
\midrule
2 & Citrinet + Parakeet                    & 0.765 & 0.277 \\
3 & W2V2 + Citrinet + Parakeet             & 0.807 & 0.295 \\
4 & Whisper + W2V2 + Citrinet + Parakeet   & \textbf{0.810} & 1.339 \\
5 & + Emformer                             & 0.780 & 2.763 \\
6 & + SpeechBrain                          & 0.800 & 2.912 \\
\bottomrule
\end{tabular}
\vspace{-0.4cm}
\end{table}
\normalsize
\vspace{-0.5cm}
\section{Conclusion}
\vspace{-0.3cm}
We presented CLEAR, a reference-free approach for estimating speech-content leakage from cross-ASR disagreement. CLEAR closely tracks speech content leakage across privacy configurations and enables practical runtime feedback through lightweight ASR ensembles and identification of potentially exposed words. These results provide a step toward speech-privacy systems that can assess and adapt their protection during deployment. 

\bibliographystyle{IEEEbib}
\bibliography{main}

\end{document}